\documentclass{aa} 

\usepackage{graphicx}
\usepackage{txfonts}
\usepackage[]{hyperref}
\usepackage{dsfont} \usepackage{mathtools}

\usepackage[capitalise]{cleveref}
\Crefname{equation}{Eq.}{Eqs.}
\Crefname{figure}{Fig.}{Figs.}
\Crefname{tabular}{Tab.}{Tabs.}
\Crefname{section}{Sect.}{Sects.}

\newcommand{\glorentz}{{u^0}}

\newcommand{\umag}{w_B}

\begin{document} 

\title{Relativistic fluid modelling of gamma-ray binaries}
\subtitle{II. Application to LS 5039}

\author{
   D. Huber \inst{1}
   \and
   R. Kissmann\inst{1}
   \and
   O. Reimer \inst{1}
}

\institute{
Institut f\"ur Astro- und Teilchenphysik \\
Leopold-Franzens-Universit\"at Innsbruck\\
6020 Innsbruck, Austria
\email{david.huber@uibk.ac.at}
}

\date{Received --; accepted --}

\abstract{
We have presented a numerical model for the non-thermal emission of gamma-ray binaries in a pulsar-wind driven scenario.
}{
We apply this model to one of the best-observed gamma-ray binaries, the LS 5039 system.
}{
The model involves a joint simulation of the pulsar- and stellar-wind interaction and the transport of electronic pairs from the pulsar wind accelerated at the emerging shock structure.
We compute the synchrotron and inverse Compton emission in a post-processing step, while consistently accounting for relativistic beaming and $\gamma \gamma$-absorption in the stellar radiation field.
}{
The stellar- and pulsar-wind interaction leads to the formation of an extended, asymmetric wind collision region developing strong shocks, turbulent mixing, and secondary shocks in the turbulent flow.
Both the structure of the collision region and the resulting particle distributions show significant orbital variation.
Next to the acceleration of particles at the bow-like pulsar wind and Coriolis shock the model naturally accounts for the reacceleration of particles at secondary shocks contributing to the emission at very-high-energy (VHE) gamma-rays.
The model successfully reproduces the main spectral features of LS 5039.
While the predicted lightcurves in the high-energy and VHE gamma-ray band are in good agreement with observations, our model still does not reproduce the X-ray to low-energy gamma-ray modulation, which we attribute to the employed magnetic field model.
}{
We successfully model the main spectral features of the observed multiband, non-thermal emission of LS 5039 and thus further substantiates a wind-driven interpretation of gamma-ray binaries.
Open issues relate to the synchrotron modulation, which might be addressed through a magnetohydrodynamic extension of our model.
}


\maketitle

\section{Introduction} \label{sec:intro}
Gamma-ray binaries are composed of an early-type, massive star in orbit with a compact object, either a neutron star or a black hole, and are distinguished from X-ray binaries by a dominant radiative output in the gamma-ray regime $>1\,$MeV \citep[see][for a review]{Dubus2013A&ARv..21...64D, Paredes2019arXiv190103624P}.
They exhibit broadband non-thermal emission, which is modulated with the orbital phase, for most systems.
\\
In the literature, two possible mechanisms are proposed to explain their non-thermal emission \citep[see e.g.][]{Mirabel2006Sci...312.1759M, Romero2007A&A...474...15R}:
A microquasar scenario, where high-energy particles are produced in relativistic jets powered by the accretion of stellar matter onto the compact object \citep[see e.g.][]{Bosch-Ramon2009IJMPD..18..347B};
or a wind-driven scenario, where the compact object is commonly assumed to be a pulsar, accelerating particles at the shocks formed in the wind collision region (WCR) through the stellar- and relativistic pulsar-wind interaction \citep[see][]{Maraschi1981MNRAS.194P...1M, Dubus2006A&A...451....9D}.
\\
In this work, we will specifically focus on the LS 5039 system, one of the most closely studied gamma-ray binaries, which constitutes a suitable testbed for different modelling approaches due to well-known orbital parameters and the wealth of available broadband data.
The LS 5039 system is composed of a massive, O-type star in a mildly eccentric ($e = 0.35$) $\sim$3.9 d orbit with a compact object \citep{Casares2005MNRAS.364..899C}.
Supported by the available data showing regular orbital modulations in X-rays \citep{Takahashi2009ApJ...697..592T}, low-energy \citep[LE,][]{Collmar2014A&A...565A..38C}, high-energy \citep[HE,][]{Abdo2009ApJ...706L..56A} and very-high-energy gamma-rays \citep[VHE,][]{Aharonian2005Sci...309..746A}, this system is widely assumed to be a representative of the wind-driven scenario, assuming the compact object to be a pulsar.
\\
The system shows correlated orbital modulations in the X-ray, LE and VHE bands, peaking at inferior conjunction (when the compact object passes in front of the star as seen by the observer) and reaching its minimum close to superior conjunction (compact object behind the star).
In contrast, the HE modulation is anti-correlated to the previously mentioned bands, peaking at superior conjunction.
\\
Most emission models for LS 5039 are purely leptonic, establishing synchrotron emission and anisotropic inverse Compton scattering on the stellar radiation field as the dominant radiative processes \citep[see e.g.][]{Zabalza2013A&A...551A..17Z, Dubus2015A&A...581A..27D, Molina2020arXiv200700543M}.
While inverse Compton scattering is most efficient at superior conjunction, i.e. when the stellar photons are back-scattered in the direction of the observer \citep[see e.g.][]{Dubus2008A&A...477..691D}, also the attenuation of the VHE flux due to the $\gamma\gamma$ pair-production process \citep[see e.g.][]{Dubus2006A&A...451....9D} is highest, since the generated radiation has to propagate through the binary system before it can escape.
Since the shocked pulsar wind retains relativistic velocities \citep[see e.g.][]{Bogovalov2008MNRAS.387...63B}, the emission reaching the observer at Earth is modulated by relativistic boosting \citep[see e.g.][]{Dubus2010A&A...516A..18D} depending on the orientation of the system.
This interplay between the generation of inverse Compton emission and its absorption by pair-production together with the phase-dependent relativistic boosting is commonly used to explain the observed anti-correlation of the HE gamma-ray emission to the X-ray and VHE gamma-ray fluxes.
\\
Due to the high $\gamma \gamma$-opacity, a one-zone model for LS 5039 with a single emitter at the stellar and pulsar wind standoff predicts non-detectable VHE fluxes at superior conjunction, whereas a faint source is still clearly detected.
This indicates that the VHE emitting region is more extended and located farther away from the star since an alternative explanation via an electromagnetic cascade initiated by the $\gamma\gamma$ pair-production turned out to be incompatible with the observed spectrum \citep[see e.g.][]{Bosch-Ramon2008A&A...489L..21B, Cerutti2010A&A...519A..81C}.
A plausible second emission region is the Coriolis shock, which is formed due to the orbital motion of the system \citep[see][]{Bosch-Ramon2011A&A...535A..20B, Bosch-Ramon2012A&A...544A..59B}.
Using an analytic approximation to describe its location, the Coriolis shock was established as a viable site for particle acceleration and the subsequent production of VHE gamma-rays \citep[see, e.g.,][]{Zabalza2013A&A...551A..17Z}.
While such an analytic two-zone model still oversimplifies the structure of the WCR, it shows that an approach that contains the precise structure of the WCR can take the extended emission region and the consequent dilution of $\gamma\gamma$-absorption into account very naturally.
To obtain a more realistic picture of the WCR, the pulsar- and stellar-wind interaction was investigated numerically using dedicated relativistic hydrodynamics (RHD) simulations \citep[see e.g.][]{Lamberts2013A&A...560A..79L, Bosch-Ramon2015A&A...577A..89B}.
These simulations were used in combination with particle transport models to build more comprehensive emission models with various degrees of complexity \citep[see e.g.][]{Dubus2015A&A...581A..27D, Molina2020arXiv200700543M}.
However, these models do not take the complex dynamic structure of the wind-interaction into account and rely on simplified simulations or analytic reductions of simulations.
\\
In \citet{Dubus2015A&A...581A..27D}, the authors solve a Fokker-Plank-type particle-transport equation for accelerated electronic pairs using results from a simplified wind-interaction simulation as steady-state background.
This allows for relativistic and anisotropic effects to be included consistently in the computation of the generated emission.
The wind interaction was simulated in a non-turbulent setup, symmetric around the binary axis, which is rescaled for different orbital phases.
While reducing the computational effort, the assumption of axisymmetry prevents orbital motion to be taken into account, implying that a Coriolis shock cannot be formed in such a simulation.
Further, with a steady-state fluid background, the impact of fluid dynamics is neglected.
However, more complex simulations \citep[e.g.][]{Bosch-Ramon2015A&A...577A..89B} have shown that neither the shock structure nor the downstream flow are stationary but depend on the orbital phase and are subject to turbulence arising in the WCR.
\\
In \citet{Huber2020-1} we presented a novel numerical model for the non-thermal emission of gamma-ray binaries in a pulsar-wind driven scenario, with which we aim to alleviate some of the approximations in previous modelling efforts while incorporating previous approaches in a self-consistent framework. 
In the presented approach the pulsar- and stellar-wind interaction is treated in a dedicated 3-dimensional relativistic hydrodynamic simulation.
The novelty of this approach consists of a simultaneous solution of the fluid dynamics and the particle transport for the electron-positron pairs, accelerated at the arising shocks.
This yields consistent particle distributions, which are used together with the fluid solutions to generate predictions for the non-thermal emission of the system, comprising synchrotron and anisotropic inverse-Compton emission with additional modulation by relativistic boosting and $\gamma \gamma$-absorption.
\\
The present paper is the second in a series of papers.
Here, we apply the model presented in \citet{Huber2020-1} to the LS 5039 system, which is a prime candidate for the investigation of fluid models of gamma-ray binaries due to the well know and less complex stellar-wind model as compared to other systems with stellar disks.
With this application, we aim to identify the relevant emission regions in three dimensions and to assess the impact of fluid dynamics on the radiative output of the system, which has not been possible with previous approaches.
\\
The paper is structured as follows:
in \cref{sec: simulation setup} we specify the setup of the simulation;
the resulting fluid solution, particle distributions and predicted emission for LS 5039 are presented in \cref{sec: results}; and a summarizing discussion is given in \cref{sec: summary}.

\section{Simulation setup}\label{sec: simulation setup}
\begin{figure}
  \centering
  \includegraphics[width=\linewidth]{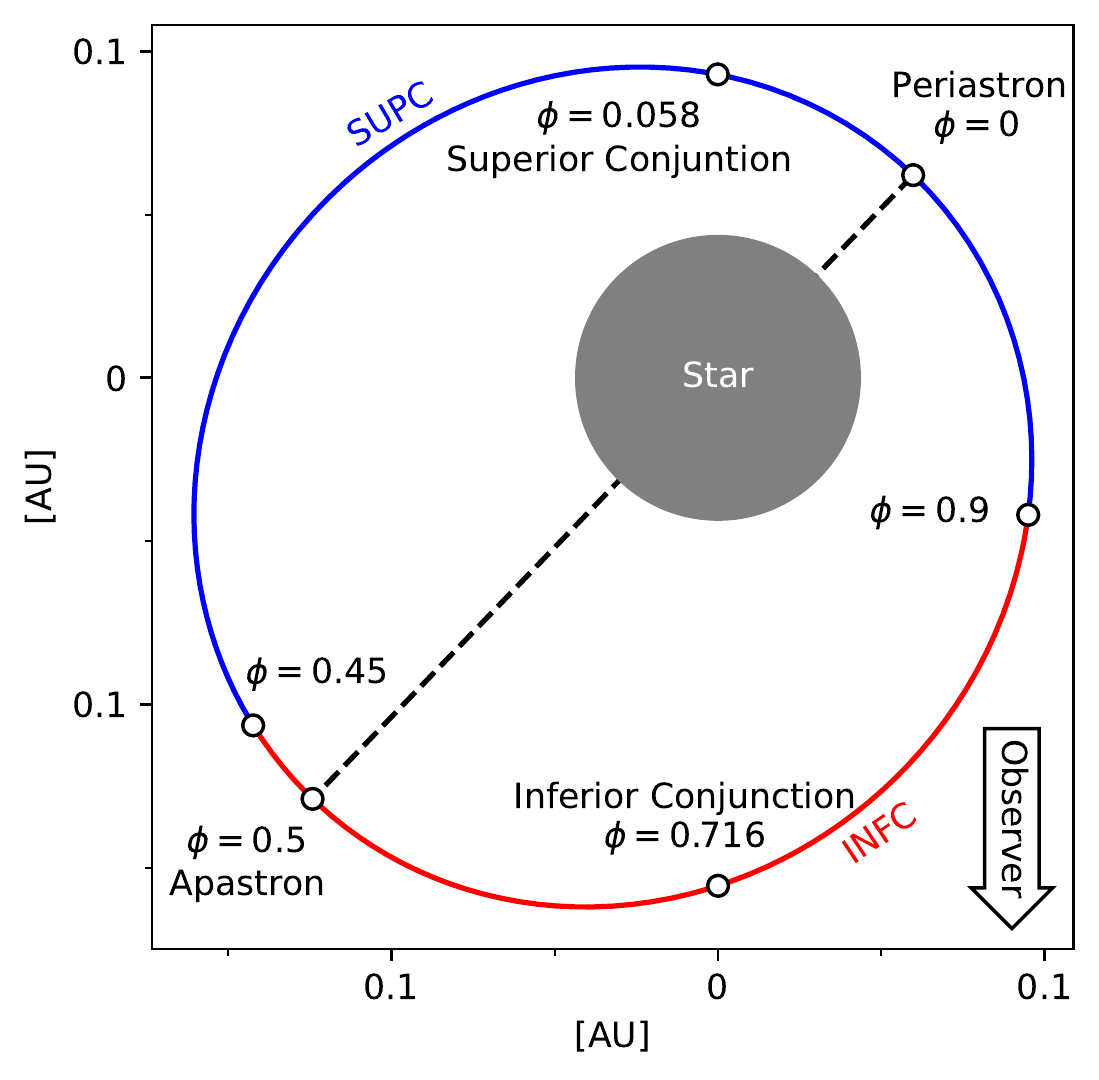}
  \caption{The LS 5039 system as determined by \cite{Casares2005MNRAS.364..899C} with an orbital period of $P_\text{orb} = 3.9\,$d and an eccentricity of $e=0.35$ assuming a stellar mass $M_\text{star} = 23\,\text{M}_\odot$ and pulsar mass $M_\text{pulsar} = 1.4\,\text{M}_\odot$.
  Periastron, apastron and the conjunctions are indicated.
  Matching the choices in \cite{Aharonian2006A&A...460..743A} the orbit is divided in INFC ranging from $0.45 < \phi < 0.9$, and SUPC with $0.9 < \phi < 1 $ and $0 < \phi < 0.45$.
  } \label{fig: ls5039 orbit}
\end{figure}
The LS 5039 system is composed of a massive, O-type star in a mildly eccentric orbit with, what we assume to be, a pulsar, despite that no pulsations have been detected, yet.
We employ the orbital solution determined by \cite{Casares2005MNRAS.364..899C}, with a period $P_\text{orb} = 3.9\,$d and eccentricity $e=0.35$.
We assume a stellar mass $M_\text{star} = 23\,\text{M}_\odot$ and a pulsar mass $M_\text{pulsar} = 1.4\,\text{M}_\odot$ \citep[similar to][]{Dubus2015A&A...581A..27D}, yielding a semi-major axis $a = 0.145\,$AU, apastron distance $d_a = 0.196\,$AU and periastron distance $d_p = 0.094\,$AU, as depicted in \cref{fig: ls5039 orbit}.
\\
To simulate the fluid of the stellar and the pulsar wind, we numerically solve the equations of special-relativistic hydrodynamics, using a reference frame co-rotating with the average angular velocity of the system $\Omega = 2 \pi / P_\text{orb}$ \citep[see][for more details]{Huber2020-1}.
The dynamical equations are solved using the \textsc{Cronos} code \citep[][]{Kissmann2018ApJS..236...53K}.
The simulation is performed on a Cartesian grid with dimensions $[-1.25,1.25]\times[-0.5,1.5]\times[-0.5,0.5]\,$AU$^3$, which is homogeneously subdivided into $640\times 512\times 256$ spatial cells.
The binary orbits in the $x-y$ plane, with its centre of mass coinciding with the coordinate origin.
\\
For the simulation of the wind interaction (presented in \cref{sec:fluid results}), we adopted a stellar mass loss rate of $\dot{M}_s = 2 \times 10^{-8} \,\text{M}_\odot\, \text{yr}^{-1}$ and a terminal velocity of $v_s = 2000 \,\text{km} \, \text{s}^{-1}$ \citep[][]{Dubus2015A&A...581A..27D}.
The speed of the pulsar wind is set to $v_p=0.99 \, \text{c}$, corresponding to a Lorentz factor of $\glorentz_p = 7.08$.
While this is smaller than conventional values $\glorentz_p \sim 10^4 - 10^6$ \citep[see e.g.][and references therin]{Khangulyan2012ApJ...752L..17K, Aharonian2012Natur.482..507A}, it is still high enough to capture the relevant relativistic effects \citep{Bosch-Ramon2012A&A...544A..59B}.
The pulsar mass loss rate $\dot{M}_p = \eta \frac{\dot{M}_s u_s}{u_p}$ is scaled with $\eta = 0.1$, yielding a corresponding pulsar spin-down luminosity
$L_\text{sd} = 7.55 \times 10^{28}\,\text{W}$.
Both, the stellar and the pulsar winds, are continuously injected within a spherical region with radius $r_\mathrm{inj} = 0.08\,a$ around each object, initially placed in a vacuum.
To account for the eccentricity of the orbit, the locations of the injection volumes are updated accordingly after every step of the simulation.
\\
The simulation is performed for approximately 1.6 orbits.
The first 0.6 orbits are used to initialize the simulation, giving the slower stellar wind time to populate the computational domain, which reaches the farthest corner of the simulation box after $t_\mathrm{crossing} \approx 0.45 P_\mathrm{orb} \approx 1.75 \,\mathrm{d}$.
The subsequent full orbit is used for the scientific analysis.
\\
For the evolution of the accelerated electron-positron pairs, we solve a Fokker-Planck-type transport equation simultaneously to the fluid interaction using an extension to the \textsc{Cronos} code \citep[][]{Huber2020-1}.
For the particles, we use 50 logarithmic bins in energy covering the range $\gamma \in [200, 4 \times 10^8]$.
At the acceleration sites, identified by a strongly compressive fluid flow $\nabla_\mu u^\mu < - 10 \, \mathrm{c}/\mathrm{AU}$, where $u^\mu$ is the fluids four-velocity, we inject two particle populations into the simulation: A powerlaw component, corresponding to accelerated, non-thermal pairs, and a Maxwellian component, corresponding to isotropised pairs from the pulsar wind \citep[see also][]{Dubus2015A&A...581A..27D}.
For brevity, we will refer to these particles only as electrons for the rest of this work.
The non-thermal electrons are injected with a spectral index $s = 1.5$ and an acceleration efficiency $\xi_\text{acc} = 2$, determining the maximum energy $\gamma_\mathrm{max}= \left( \frac{3}{4\pi} \frac{\mu_0 e c}{\xi_\text{acc} \sigma_T B'} \right)^{1/2}$ reached by balancing diffusive shock acceleration and synchrotron losses with the magnetic field strength $B'$.
We assume a fraction $\zeta_e^\text{PL} = 0.45$ of the local internal fluid energy and a fraction $\zeta_n^\text{PL} = 4 \times 10^{-3}$ of the number of available electrons from the pulsar wind to be injected as non-thermal component.
The remaining electrons are injected as a Maxwellian component.
Since the pulsar-wind Lorentz factor used in our simulation might be higher in reality, the density of the pair plasma will be overestimated in our simulation.
To account for this, we decrease the number of available electrons considered in the injection by a factor $\zeta_\rho = 5.5 \times 10^{-4}$.
\\
In our simulation, we do not evolve the magnetic field explicitly, which makes assumptions on its spatial dependence necessary.
Therefore, we conduct the simulations under the assumption that the magnetic-field energy density in the fluid frame is proportional to the internal energy density of the fluid $\frac{B'^2}{2\mu_0} = \zeta_b \frac{p}{\Gamma-1}$ \citep[in analogy to][]{Barkov2018MNRAS.479.1320B} with $\zeta_b = 0.5$.
This approximation is similar to the one used in \citet{Dubus2015A&A...581A..27D} at the location of the shocks.
Although the evolution of the magnetic might differ in the shocked downstream flow, this approximation is expected to yield good values at the shocks, which is most relevant for the particle injection.
For the stellar companion, we assume a luminosity $L_\star = 1.8 \times 10^5\, \text{L}_\odot$, temperature $T_\star = 39000\,\text{K}$ and resulting radius $R_\star = 9.3 \, \text{R}_\odot$ \citep{Casares2005MNRAS.364..899C}. 
\section{Results}\label{sec: results}
\subsection{Fluid structure}\label{sec:fluid results}
\begin{figure*}
  \centering
  \includegraphics[width=\linewidth]{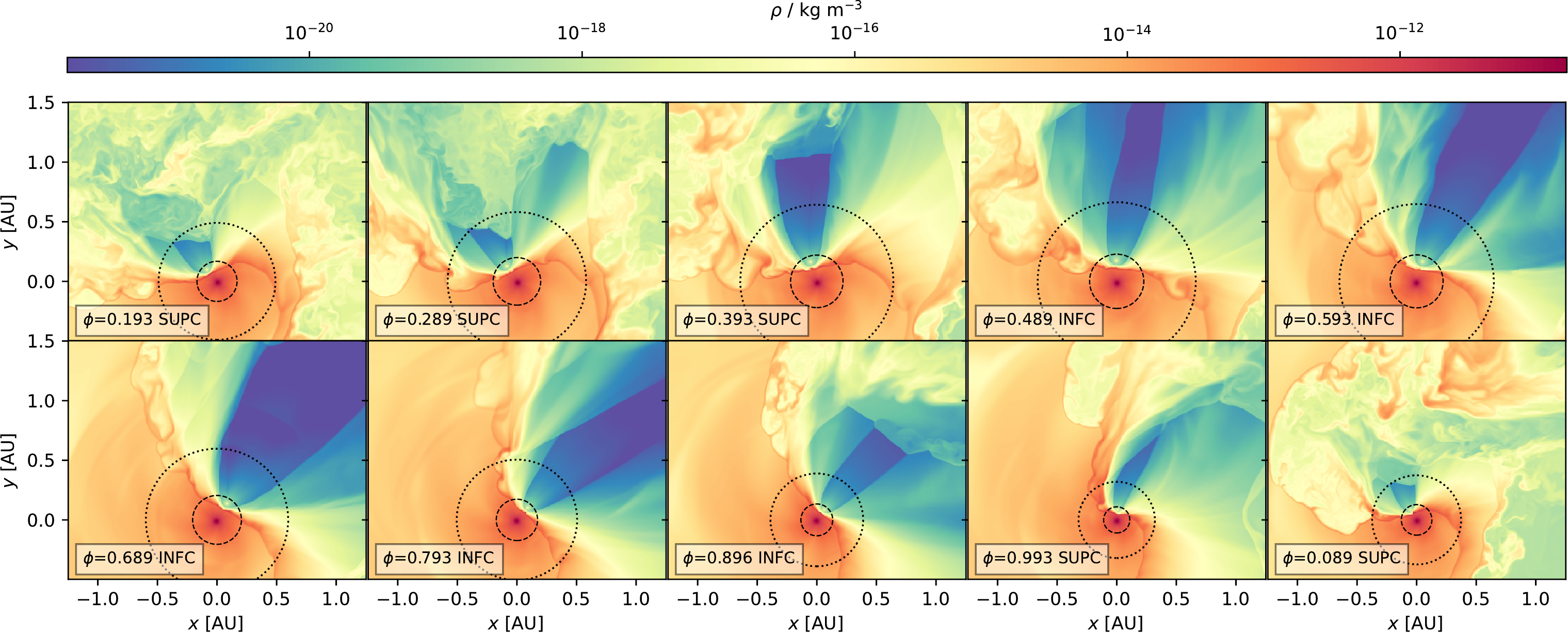}
  \caption{
    The fluid's mass density in the orbital plane for different orbital phases, indicated by the inset annotations.
    An extended WCR is formed due to the interaction of the pulsar (blue) and stellar (orange) wind.
    The reference frame is corotating with counter-clockwise orbital motion.
    The labels (INFC/SUPC) correspond to respective parts of the orbit as defined in \cref{fig: ls5039 orbit}.
    The boundaries of the inner, middle and far region (further discussed in the text) are indicated by the dashed and dotted lines, respectively.
  } \label{fig: results fluid density}
\end{figure*}In \cref{fig: results fluid density} we show the resulting fluid mass-density for different orbital phases.
An extended wind collision region (WCR) is formed from the stellar- and pulsar-wind interaction, creating a bow-like shock (hereafter bow shock) on the pulsar side.
The overall shape of the WCR is highly asymmetric and bends in the direction opposite to the orbital motion due to the Coriolis force acting on the rotating system.
The additional force leads to a termination of the pulsar wind on the far side of the system forming another shock at the Coriolis turnover (hereafter Coriolis shock).
\\
In the downstream region of the Coriolis shock, shocked pulsar wind is turbulently mixed with shocked stellar-wind material, slowing the flow due to increased mass load (see also \cref{fig: results fluid four speed}).
Many secondary shocks are formed in the turbulent motion of the fluid, which can re-energize already cooled particles accelerated at other shocks \citep{Bosch-Ramon2012A&A...544A..59B}.
In the wings of the WCR, the pulsar wind shocked at the bow is reaccelerated towards its initial Lorentz factor \citep[see also][]{Bogovalov2008MNRAS.387...63B} until it reaches the turbulently mixed fluid behind the Coriolis shock creating an additional shock (hereafter reflected shock in analogy to \citet{Dubus2015A&A...581A..27D}) extending diagonally into the region behind the pulsar (clearly visible at $\phi \simeq 0.3$ in \cref{fig: results fluid density} and also \cref{fig: results fluid thermal energy}).
Features visible in our simulation are in qualitative agreement with previous works \citep[see e.g.]{Bosch-Ramon2015A&A...577A..89B}, validating our fluid model.
\\
While the Coriolis shock location is in agreement with previous works at periastron, its distance from the pulsar becomes too large and leaves the computational domain around apastron.
This suggests that the simulation box does not capture enough of the leading edge of the WCR, which therefore cannot sweep up enough dense stellar-wind material to build up the required pressure to terminate the pulsar wind.
\\
The structure of the WCR is strongly influenced by turbulence developing at the contact discontinuity in the WCR.
Next to the Kelvin-Helmholtz (KH) instabilities triggered by the large velocity shear, the arising turbulence is further increased by Rayleigh-Taylor (RT) and Richtmeyer-Meshkov (RM) instabilities \citep{Bosch-Ramon2012A&A...544A..59B}.
This dominantly occurs at the leading edge of the WCR (in front of the pulsar with respect to the counter-clockwise orbital motion), because there the high-density, shocked stellar wind is pressed against the low-density, shocked pulsar wind by the Coriolis force \citep[see also][]{Bosch-Ramon2012A&A...544A..59B}.
We found that the formation of the Coriolis shock is influenced by the arising turbulence (see e.g. $\phi \simeq 0.8$ in \cref{fig: results fluid density}), that can cause orbit-to-orbit variations in the WCR structure.
\\
The development of KH instabilities is strongly suppressed for highly relativistic flows \citep[see e.g.][]{Perucho2004A&A...427..431P,Bodo2004PhRvE..70c6304B}.
Fluctuations can therefore only efficiently grow in the vicinity of the wind standoff, where the shocked pulsar wind is sufficiently slow, having a speed of $\lesssim c/3$, before it is reaccelerated.
The timescale for advection out of this region therefore critically determines the largest modes that can be excited in the system, since the growth-rate for KH instabilities $\Gamma_\mathrm{KH} \propto \lambda^{-1}$ decreases with larger wavelengths \citep[see][]{Bodo2004PhRvE..70c6304B, Lamberts2013A&A...560A..79L}.
The maximum wavelength of fluctuations that can be excited in the presented system is on the order of $\lambda_\mathrm{max} \sim 0.1\,$AU (assuming a region extension of $\sim 0.3\,$AU and an average advection velocity of $\sim 0.5\,$c).
\\
The pulsar-wind Lorentz factor used in our simulation is lower than typically assumed values $\sim 10^4 - 10^6$, due to numerical limitations.
A higher pulsar wind Lorentz factor will yield a faster reacceleration \citep[][]{Bosch-Ramon2012A&A...544A..59B} when the shocked flow is expanding in the wings and might therefore reduce the advection timescale and shift the maximum wavelength of growing fluctuations to smaller values.
However, \citet{Bogovalov2008MNRAS.387...63B} have found that even in the ultra-relativistic limit, the shocked pulsar-wind flow remains slow near the head of the bow-shock up to scales of the orbital separation for our choice of $\eta$.
This is mainly a consequence of the bow-like geometry of the WCR, causing the upstream incident angle with the shock normal to be small over a wide part of the head of the bow shock leading to a strongly slowed downstream flow.
Further, the flow cannot expand rapidly there, keeping the reacceleration at a moderate level.
The reacceleration only becomes more important farther out, where the opening angels of the pulsar-wind shock and the contact discontinuity approach their asymptotic values.
The extent of the unstable region hence cannot go below the scales on the order of the orbital separation regardless of the pulsar-wind Lorentz factor, thus only slightly changing the estimates from above.
Even in the ultra-relativistic limit, KH instabilities are therefore still expected to grow and the overall structure of the WCR should not change significantly.
\\
For our setup, we have found in numerical tests \citep[simulating the decay of shear flows as described by][]{Ryu1994ApJ...422..269R} that damping by numerical viscosity $\Gamma_\mathrm{damp} \propto \lambda^{-3}$ dominates over the KH growth-rate \citep[estimated from][]{Bodo2004PhRvE..70c6304B} for short-wavelength fluctuations with $\lambda \lesssim 10^{-2}\,$AU.
The chosen resolution, therefore, allows the growth of fluctuations in the range $10^{-2} \, $AU$ \lesssim \lambda \lesssim 10^{-1}\,$ AU (as seen in \cref{fig: results fluid density}), where the latter will have the biggest impact on the WCR.
This also implies that future higher-resolution simulations will extend the range of turbulence to smaller spatial scales for which the driving by KH instabilities becomes more efficient because $\Gamma_\mathrm{KH} \propto \lambda^{-1}$.
\\
In our simulation, we do not consider any feedback of the accelerated particles onto the fluid dynamics.
Because the non-thermal electrons contribute a significant fraction to the overall energy density, their cooling will result in a decrease of the fluid's pressure supporting the WCR.
Consequently, when taking such feedback into account, the WCR will be less extended and might thus be more susceptible to thin-shell instabilities \citep[as pointed out by][]{Dubus2015A&A...581A..27D} as it was seen in the case of colliding-wind binaries by \citet{Reitberger2017ApJ...847...40R}.
The consideration of such effects is left for future work.

 \subsection{Particle distribution}\label{sec:particle results}
\begin{figure}
  \centering
  \includegraphics[width=\linewidth]{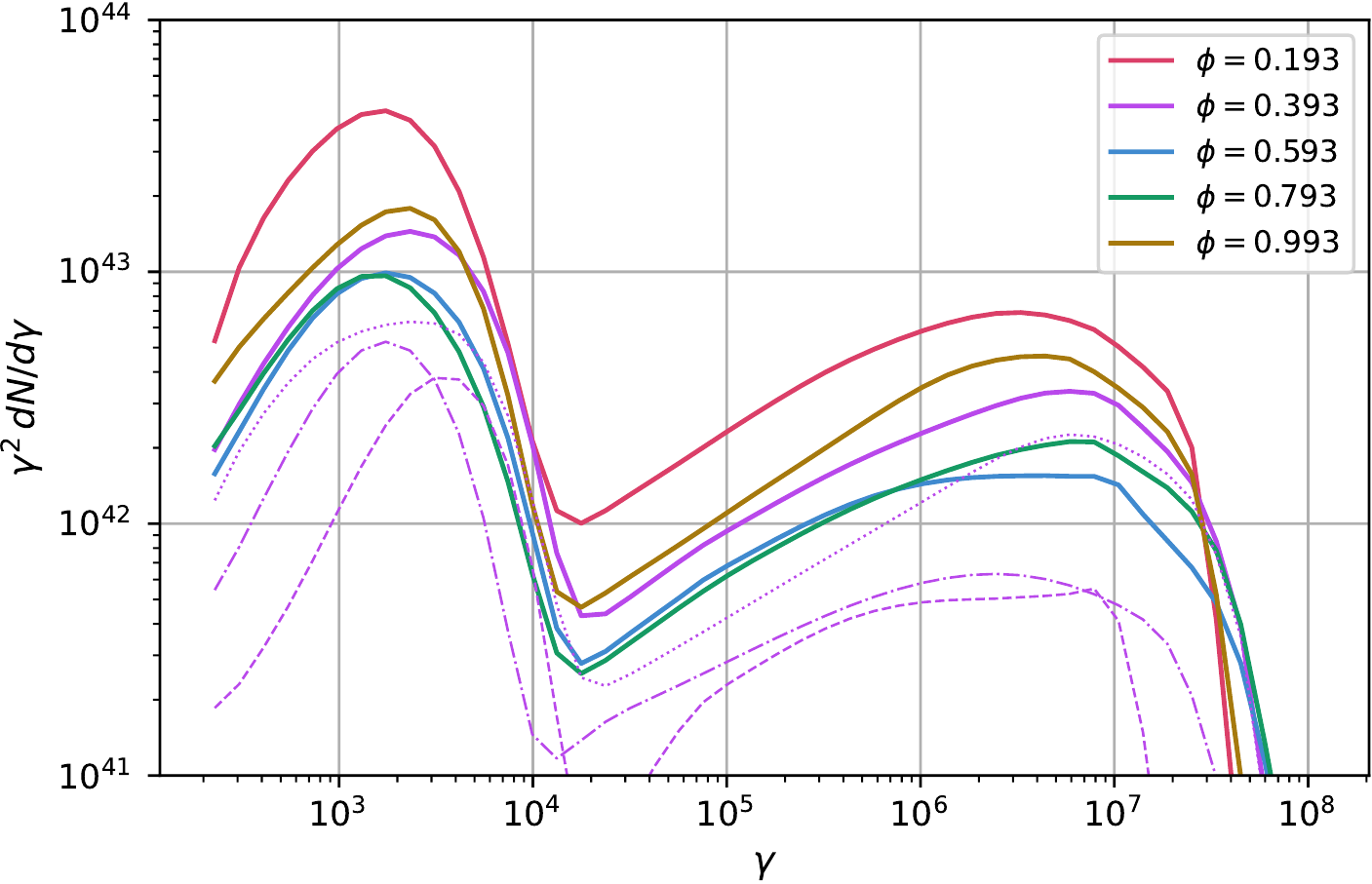}
  \caption{
    Spectral energy distribution of electrons integrated over the computational domain for different orbital phases (colour-coded, solid).
    For $\phi = 0.393$, the contributions of the inner (dashed), middle (dash-dotted), and outer (dotted) region are shown, respectively, as an example.
  } \label{fig: particle spectrum}
\end{figure}

\begin{figure}
  \centering
  \includegraphics[width=\linewidth]{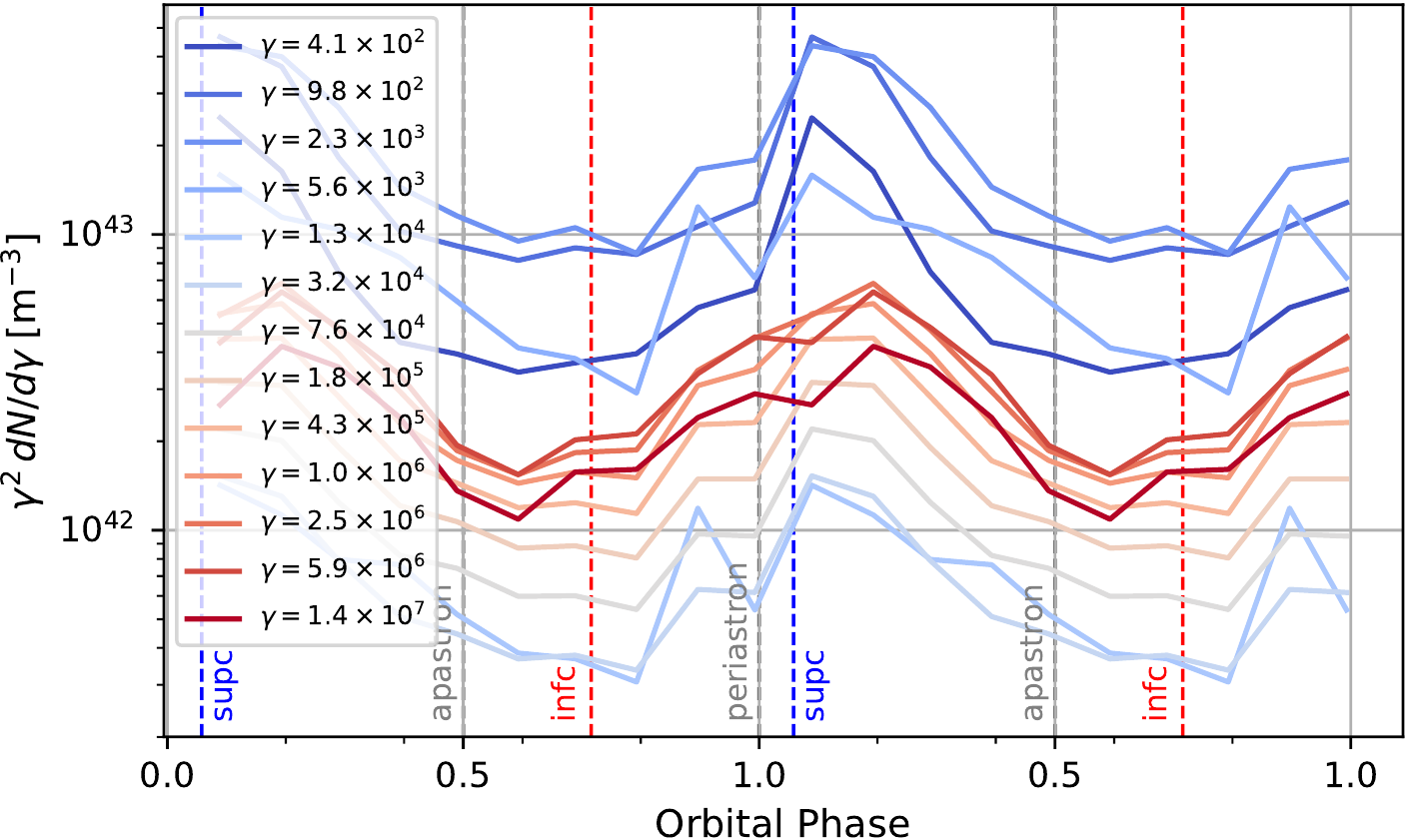}
  \caption{
    Temporal evolution of the spectral energy distribution of energetic electrons integrated over the computational domain for selected energies (colour-coded). 
    For better visualisation, we show the same data for two orbits.
  } \label{fig: particle timecurve}
\end{figure}

Since the timescale for the particle populations to reach a quasi-steady state (by reaching the limits of the computational domain and/or by cooling) is orders of magnitudes smaller than the orbital timescale we do not perform a particle transport simulation for the full timespan of the fluid simulation.
Instead, to reduce the computational effort, we restart a particle transport simulation on previously obtained fluid solutions for the 10 chosen orbital phases shown in \cref{fig: results fluid density}.
We perform these simulations for $t=1.11\,$h allowing the injected electrons to populate the system.
\\
To simulate the energetic evolution of the electrons, we employ the semi-Lagrangian scheme described in \citet{Huber2020-1}.
For this, we use the same timestep as for the preceding hydrodynamic step that is mainly determined by the speed of the unshocked pulsar-wind and the spatial resolution yielding typical values of $\Delta t \sim 0.8\,\mathrm{s}$.
Due to the semi-Lagrangian nature of the scheme, an additional reduction of the timestep depending on the energy resolution is not required to maintain stability.
\\
\begin{figure*}
  \centering
  \includegraphics[width=\linewidth]{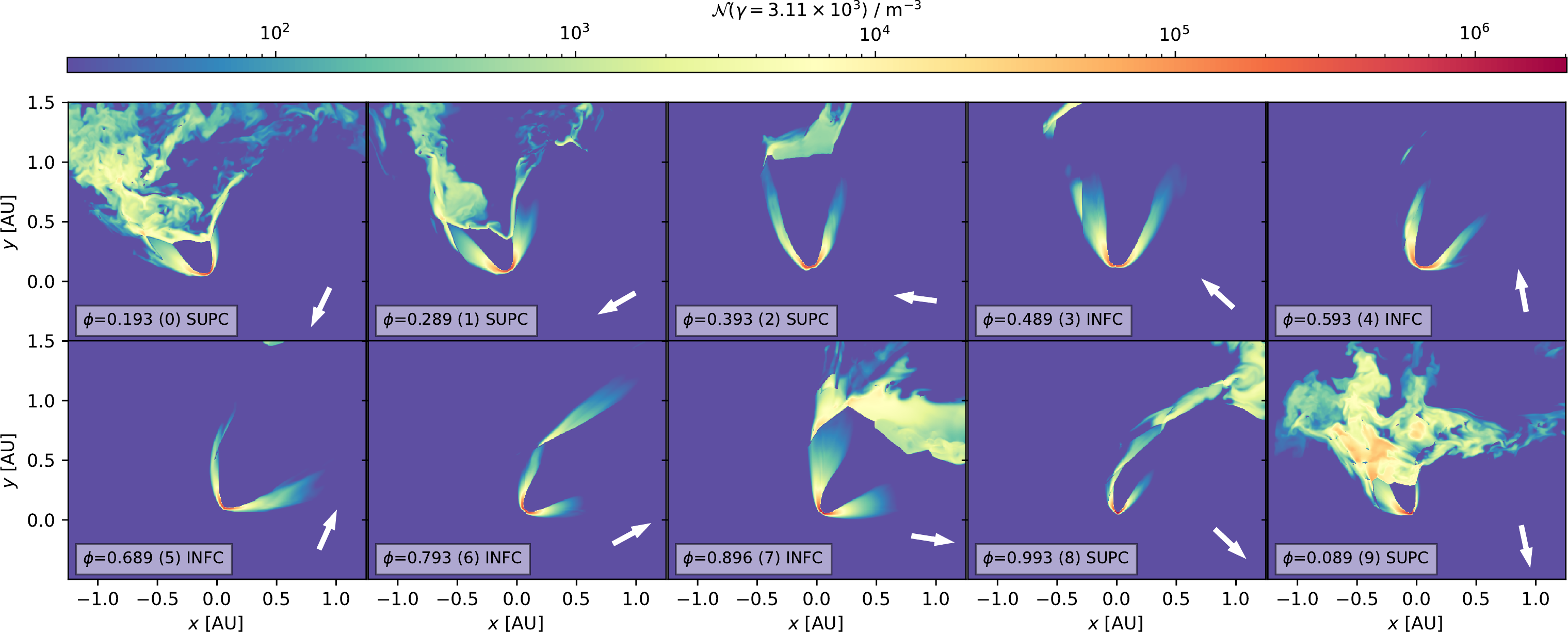}
  \caption{
    Differential particle number density for electrons with $\gamma = 3.11 \times 10^3$ in analogy to \cref{fig: results fluid density}.
    The arrows indicate the projected direction to an observer at Earth.
    For clarity, we show only 5 orders of magnitude below the highest value - the blue regions correspond to those with lower values.
  } \label{fig: particle density low}
\end{figure*}\begin{figure*}
  \centering
  \includegraphics[width=\linewidth]{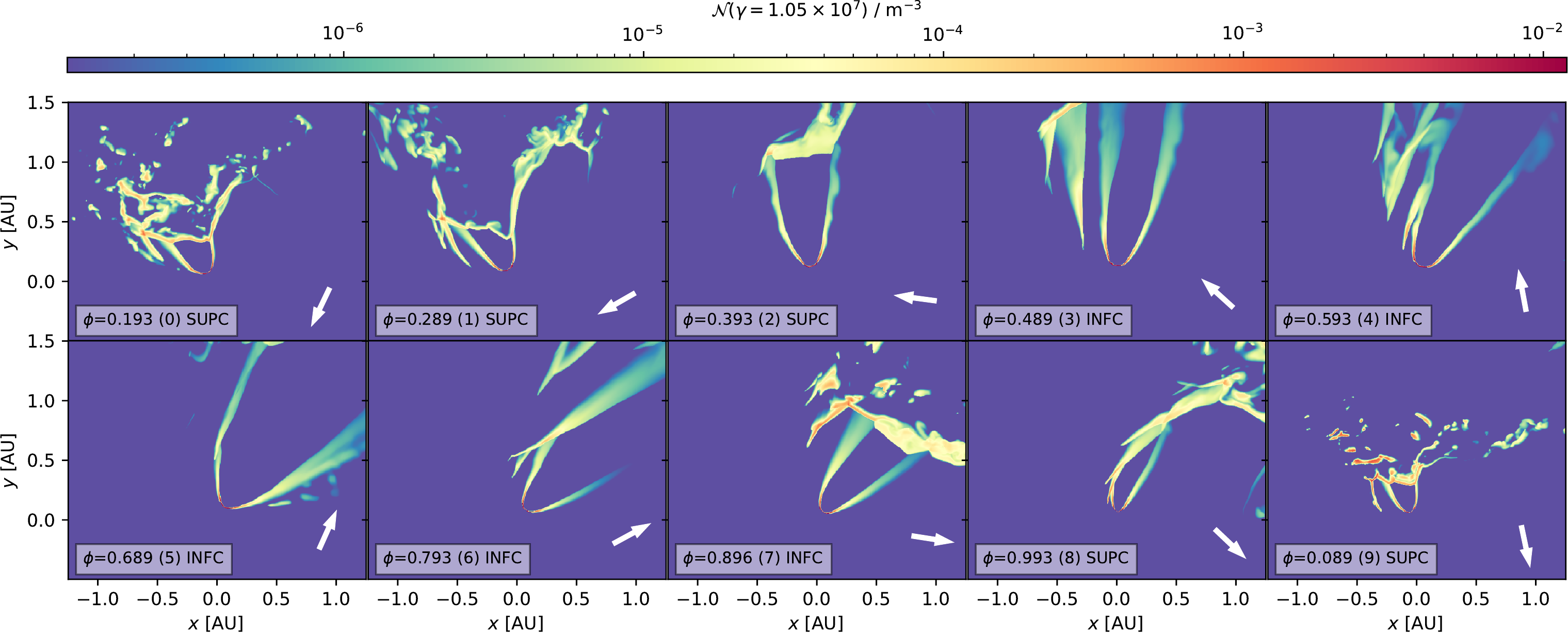}
  \caption{
    Same as \cref{fig: particle density low}, but for $\gamma = 1.05 \times 10^7$.
  } \label{fig: particle density high}
\end{figure*}In \cref{fig: particle density low} and \cref{fig: particle density high} we present the resulting electron distributions for different orbital phases at two energies, dominantly populated by Maxwellian and power-law electrons, respectively.
The relevant shock structures are imprinted in the electron distributions.
This is especially visible for particles at higher energies, which are quickly cooled through synchrotron and inverse Compton losses.
Due to increased losses, they remain much more confined to their injection sites.
Lower-energetic electrons, in comparison, dominantly lose energy via adiabatic losses and are therefore cooled on longer timescales, populating a much larger spatial domain.
Next to the bow and Coriolis shocks also the reflected shock and the countless secondary shocks in the turbulent downstream region behind the pulsar are manifestly visible.
In our simulation, radiative losses are highest at the wind standoff, which limits the maximum electron energy reached in the acceleration and the length-scales over which non-thermal electrons are cooled afterwards.
At the Coriolis shock, both the magnetic and stellar-radiation field are weaker in comparison and the highest particle energies of the simulation are reached (see \cref{fig: particle spectrum}).
\\
In \cref{fig: particle spectrum} we show the spectral energy distribution of the electrons integrated over the computational volume.
The contributions of the Maxwellian and the power-law electrons can be easily distinguished as they dominate the energy range below and above $\gamma \sim 2 \times 10^4$, respectively.
It is apparent that not only the normalisation but also the shape of the spectrum is a function of the orbital phase, considering the location of the peak in the Maxwellian electrons, the break in the power-law tail and the maximum energies reached.
Furthermore, it shows that a higher electron density is reached after periastron, which is not intuitive at first glance.
\\
This becomes more apparent in \cref{fig: particle timecurve}, where we show the temporal evolution of the electron distribution integrated over the computational volume at selected energies.
Apparently, the number of electrons in the computational volume lags behind the orbit, i.e. the minimum amount of electrons is reached after apastron and the maximum after periastron.
\\
This delay is caused by the inertia of the fluid.
The WCR needs a certain time to build up or dissipate its pressure, which critically determines the amount of electrons accelerated at the shocks.
This phenomenon can only be seen in approaches that treat the particle transport together with the underlying fluid.
The varying size of the shocks counteracts parts of this effect by shrinking and enlarging the acceleration sites at periastron and apastron, respectively, which is, however, not dominant in our simulation.
The power injected at the shocks reaches its maximum at $\phi \simeq 0.2$ in our simulation, suggesting an expected peak in the integrated electron distribution with the same delay.
\\
We find, however, that this delay is energy-dependent.
While the number of higher-energetic electrons peaks as expected around $\phi \simeq 0.2$, the number of lower-energetic ones peaks earlier at $\phi \simeq 0.1$ (see \cref{fig: particle timecurve}).
This behaviour originates in the turbulent flow behind the pulsar formed earlier around the periastron passage, which leads to a pile-up of electrons at lower energies yielding an earlier peak in their evolution.
This effect is not relevant for electrons at higher energies since they are cooled much faster through radiative losses.
Consequently, these particles are more confined to the shocks preventing a pile-up and leaving their number density more directly affected by the acceleration process.
\\
We treat both the spectral slope and the acceleration efficiency as free model parameters since the specifics for particle acceleration in gamma-ray binaries have not been firmly identified, yet.
In this phenomenological approach, we treat all acceleration sites the same, which might not be the case in reality.
For example, in the case of diffusive shock acceleration, both the spectral index and the acceleration timescale depend on the shock's compression ratio, magnetic obliquity, and up- and downstream fluid velocities \citep[see e.g.][]{Keshet2005PhRvL..94k1102K, Takamoto2015ApJ...809...29T}.
In future efforts, the current model can be extended to investigate different acceleration mechanisms.  
 \subsection{Emission}\label{sec:emission results}
\begin{figure*}
  \centering
  \includegraphics[width=\linewidth]{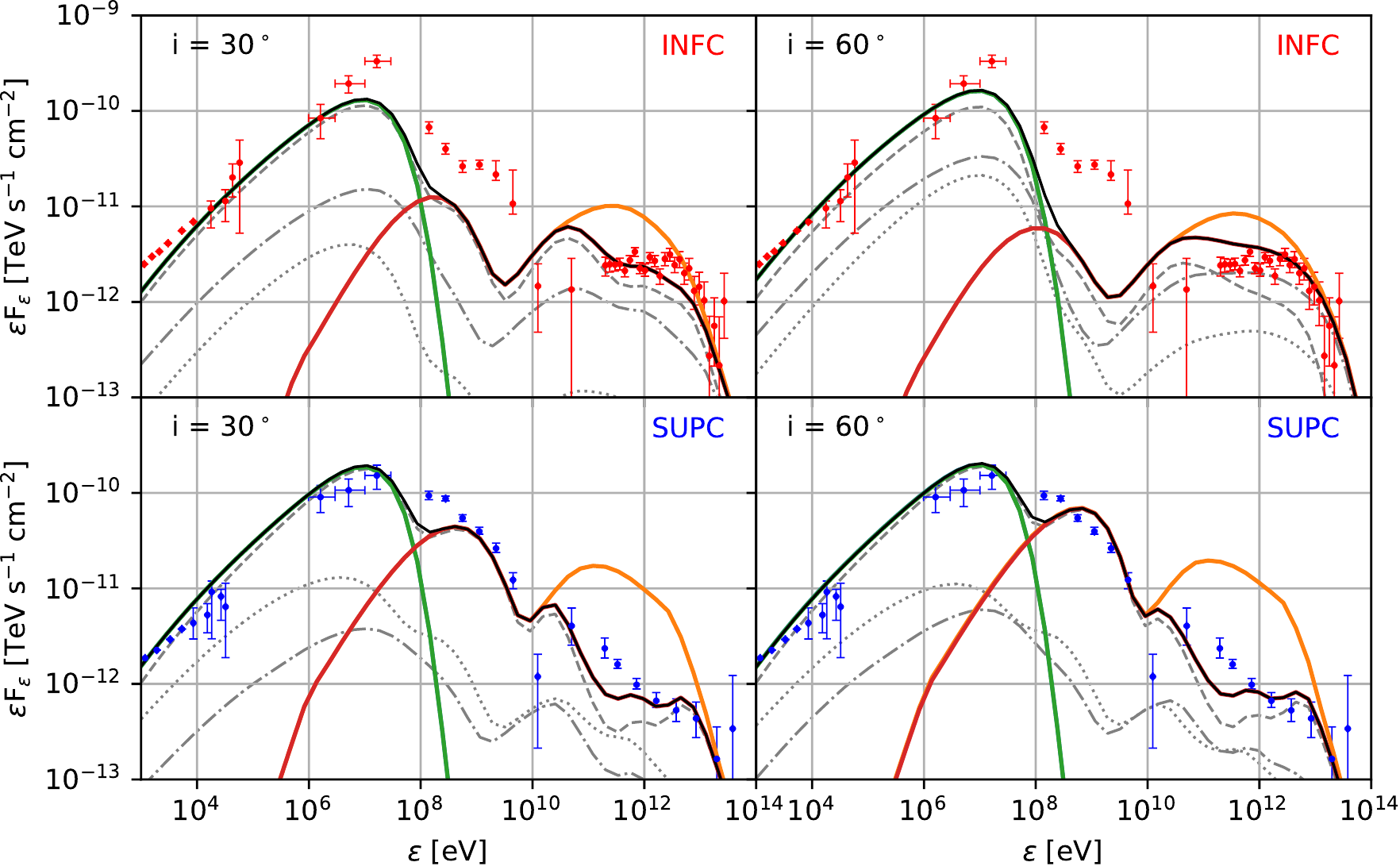}
  \caption{
    Spectral energy distribution of the emission predicted by our model for LS 5039 for different parts of the orbit and inclinations of the system. 
    Different radiative processes are colour-coded: synchrotron (green), inverse Compton (orange), inverse Compton attenuated by $\gamma \gamma$-absorption (red) and total (black).
    The contributions to the total emission from the inner (dashed), middle (dashed-dotted) and far (dotted) region are shown separately (for a description of the regions see the main text). 
    The model predictions are shown together with observations in soft X-rays \citep[][]{Takahashi2009ApJ...697..592T}, LE \citep[][]{Collmar2014A&A...565A..38C}, HE \citep[][]{Hadasch2012ApJ...749...54H} and VHE \citep[][]{Aharonian2006A&A...460..743A} gamma-rays.
    Results are shown for different parts of the orbit as defined in \cref{fig: ls5039 orbit} with INFC on the left and SUPC on the right column.
    In the top row results are shown for an inclination $i=30^\circ$ of the orbital plane, while in the second row, the results for an inclination $i=60^\circ$ are presented.
  } \label{fig: emission averaged sed}
\end{figure*}\begin{figure*}
  \centering
  \includegraphics[width=\linewidth]{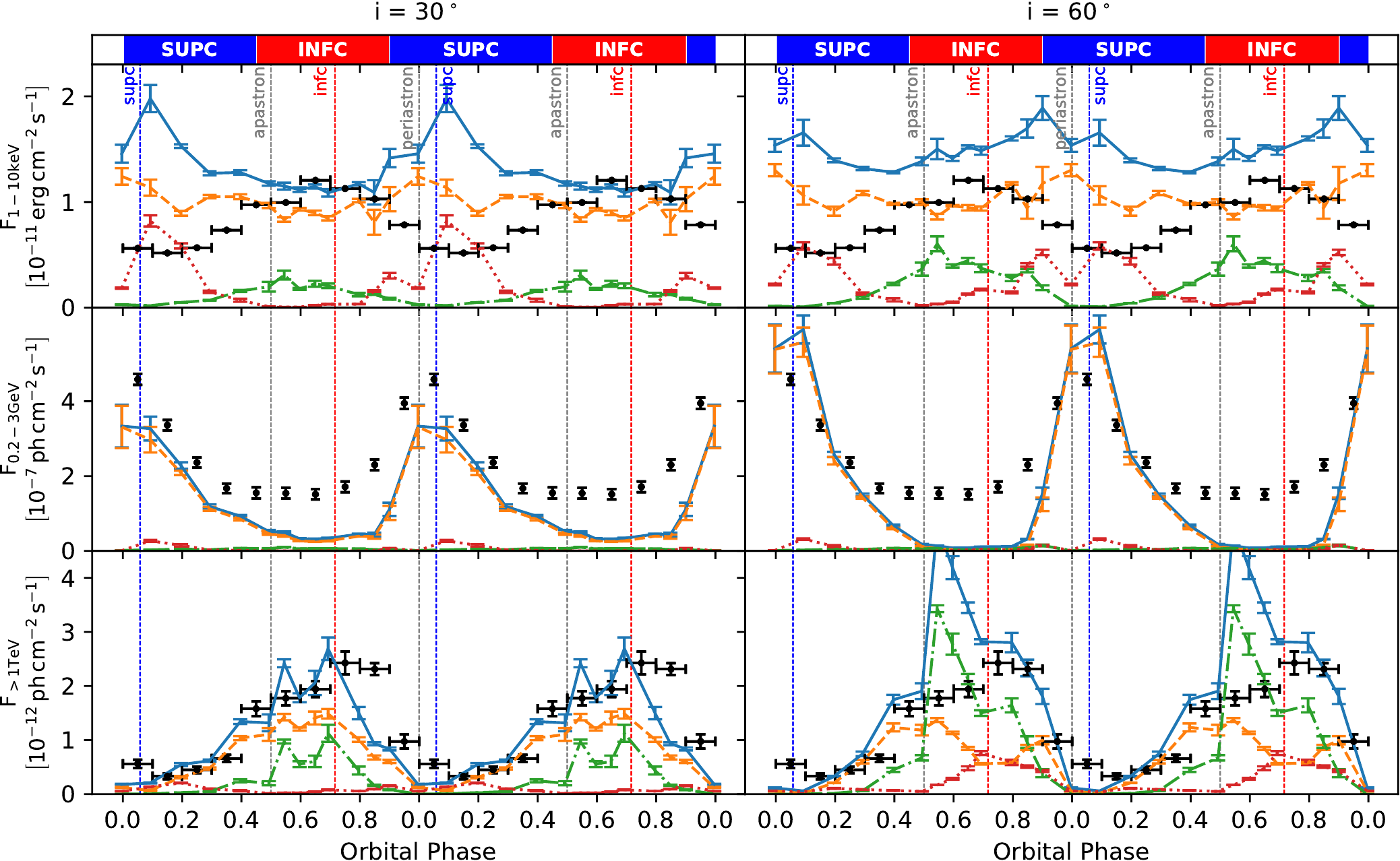}
  \caption{
    Emission lightcurves predicted by our model for LS 5039 for different energy bands and inclinations of the system.
    The contributions from the inner (orange, dashed), middle (green, dashed-dotted) and far (red, dotted) region together with their sum (blue, solid) are shown (for a description of the regions see the main text). 
    For better visualisation, we show the same data for two orbits.
    The results are shown for different energy bands together with observations in indivdual rows (from top to bottom): Soft X-rays \citep[$1-10\,$keV,][]{Takahashi2009ApJ...697..592T}, HE \citep[$0.2-3\,$GeV,][]{Hadasch2012ApJ...749...54H} and VHE \citep[$>1\,$TeV,][]{Aharonian2006A&A...460..743A} gamma-rays.
    In the left column results are shown for an inclination $i=30^\circ$ of the orbital plane, while in the right column, the results for an inclination $i=60^\circ$ are presented.
    In addition to the orbital phases described in the main text, we show 3 more lightcurve points at $\phi \sim 0.55, 0.65, 0.85$.
    The error bars indicate the variability introduced by turbulence (see \cref{sec: variability}).
  }\label{fig: emission lightcurves}
\end{figure*}In this section, we present the simulation results for the non-thermal emission spectrum and lightcurves of LS 5039.
They are produced in a post-processing step, using the previously obtained particle distributions (see \cref{sec:particle results}) and fluid solutions (see \cref{sec:fluid results}).
We, therefore, generate direct predictions on the initially spelt-out set of model-parameters (see \cref{sec: simulation setup}).
\\
For the computation of the inverse Compton emission, we treat the stellar photon field as monochromatic, whereas for the $\gamma \gamma$-absorption a full blackbody spectrum is considered.
In both cases, the source is treated as an extended sphere.
We compute the emission for two different system inclinations $i=30^\circ$ and $i=60^\circ$.
\\
In our approach, it is no longer straightforward to separate the emission produced through particles accelerated at the different shocks as it was possible in previous works \citep[e.g.][]{Dubus2015A&A...581A..27D}, since the downstream flows are turbulently mixed which was previously neglected.
To obtain a basic understanding of the location dependence of the emission we subdivide the computational volume into three regions (henceforth called inner, middle, and far region, respectively), which are separated at distances of $1.2\, d$ and $3.5\,d$ from the binaries centre of mass, where $d$ is the time-dependent orbital distance (see \cref{fig: results fluid density}).
In \cref{sec: emission spectrum} we present the predicted emission spectra followed by a more detailed discussion for individual energy-bands in \cref{sec: emission X-ray}, \cref{sec: emission he} and \cref{sec: emission vhe}.
A projection map of the emission is discussed in \cref{sec: emission map}.

\subsubsection{Emission spectrum}\label{sec: emission spectrum}
In \cref{fig: emission averaged sed} we show the resulting spectral energy distributions of photons emitted for two different inclinations of the system, integrated over different orbital phases:
for comparison to observations, the orbit has been split into two parts, INFC and SUPC as defined in \citet{Aharonian2006A&A...460..743A} (see \cref{fig: ls5039 orbit}).
To compute the average spectra we integrate the simulated ones using piece-wise linear interpolations in time.
\\
The spectral features of the particle population leave clear imprints in the emission spectrum.
Ranging from X-rays up to LE gamma-rays the spectrum is dominated by synchrotron emission generated by electrons at the power-law tail of the spectrum, dominantly produced in the inner region.
The same population of electrons is responsible for the inverse Compton scattering of stellar UV photons to the VHE gamma-ray regime, although, the relative contributions of the individual regions to the VHE emission is not constant but varies along the orbit.
The emission in the HE gamma-ray regime is also produced through the inverse Compton process, however, by the lower-energetic Maxwellian electrons in the inner region of the system.

\subsubsection{X-Rays and LE gamma-rays}\label{sec: emission X-ray}
For the chosen set of parameters, the synchrotron emission predicted by our model is able to explain the observed X-ray to LE gamma-rays emission spectrum (see \cref{fig: emission averaged sed}).
This was not possible in previous studies \citep[][]{Dubus2015A&A...581A..27D,Molina2020arXiv200700543M}.
Synchrotron emission dominates up to $\sim 10\,$MeV, with a spectral cutoff approximately constant over the orbit.
The electrons emitting at the LE cutoff are at the highest end of the power-law tail, accelerated at the apex of the bow shock.
These particles are strongly confined to shocks and, thus, rely on our injection model and the conditions directly at the shock.
This means that the synchrotron cutoff directly depends on two factors: the maximum electron energy (see \cref{sec: simulation setup}) and the maximum photon energy produced through the synchrotron process.
Both quantities only depend on the magnetic field, however, counteracting each other leading to a constant synchrotron cutoff.
\\
While the model predictions closely resemble the X-ray and LE gamma-ray observations at INFC, the X-ray flux is over-predicted at SUPC.
This deviation is also manifest in the X-ray lightcurve (see the first row of \cref{fig: emission lightcurves}).
Instead of the observed correlation with VHE gamma-rays, the predicted X-ray lightcurve is correlated with the HE gamma-ray flux.
This is caused by the still simplistic magnetic field description employed in our model:
Since the magnetic field strength directly scales with the fluid pressure, its maximum at the bow shock is reached shortly after periastron.
In addition, also electron acceleration is increased in this part of the orbit. 
The combined effects lead to a modulation of the synchrotron emissivity that dominates over the one introduced by relativistic boosting, which was argued to be the dominant source of variability in this band \citep[][]{Dubus2015A&A...581A..27D}.
These problems, therefore, highlight the necessity for a more sophisticated magnetic field description in future modelling.
\\
We also investigated the effects of different magnetic field models, such as a passively advected magnetic energy density $\umag$ injected together with the pulsar wind and/or a magnetic field aligned with the fluid bulk motion.
The injected magnetic energy density is scaled with the injected kinetic energy density as $\umag = \sigma \frac{\dot{M}_p c^2 u^0_p}{4 \pi r^2 u_p}$ with the pulsar wind magnetisation fraction $\sigma$, yielding the magnetic field in the laboratory frame $B = \sqrt{2 \mu_0 \umag}$.
We found little difference in the resulting emission spectra when using these alternative models for similar magnetic field strengths, still yielding disagreement with the predicted X-ray lightcurve.
\\
For larger inclination angles (see the right column in \cref{fig: emission lightcurves}), relativistic boosting becomes more important.
This decreases the X-ray flux at superior conjunction while increasing it at inferior conjunction, which is better resembling the features of the observed lightcurve.

\subsubsection{HE gamma-rays}\label{sec: emission he}
While the overall spectral shape and the flux of the predicted HE emission are in good agreement with observations at SUPC, the predictions prove to be problematic at INFC (see \cref{fig: emission averaged sed}).
There, the predicted HE flux drops by approximately one order of magnitude and the spectrum is shifted towards lower energies leaving the fluxes underpredicted.
These changes in the flux are mainly caused by the anisotropic inverse Compton scattering cross-section, which grows with scattering angle.
At phases around superior conjunction, stellar photons have to be scattered by a larger angle to reach the observer as compared to inferior conjunction.
Consequently, the highest and lowest inverse Compton fluxes relate to these phases, respectively.
The anisotropic scattering further causes the shift in energy, yielding higher scattered photon energies for larger scattering angles.
Both effects can also be seen by comparing the different inclination angles (e.g. see \cref{fig: emission averaged sed}).
\\
An additional modulation is generated by the changing stellar seed radiation field density induced by the varying distance of the bow shock apex to the star, with the maximum and minimum stellar photon density at periastron and apastron, respectively.
Due to the geometry of the LS 5039 orbit, the periastron passage occurs very close to superior conjunction leading this modulation to add to the former one.
\\
The underestimation of the HE gamma-ray flux at INFC suggests a missing component in our broadband emission model, which could arise from the magnetospheric emission of the pulsar as suggested e.g. by \citet{Takata2014ApJ...790...18T}.
This process can naturally account for the phase-independent spectral cutoff of the HE emission at energies of a few GeV.
To accommodate this emission within our modelling one has to find a new set of parameters for the Maxwellian electron distribution to avoid overestimation of the flux at SUPC.
\\
The discrepancy is also apparent in the predicted HE lightcurves (see the second row in \cref{fig: emission lightcurves}), showing a phase-independent underestimation with respect to the observations.
Disregarding the constant underprediction, the lightcurve for an inclination of $i=30^\circ$ is in good agreement with observations.
For the higher inclination angle, the inverse Compton anisotropy leads to a steep rise in emission at superior conjunction, while further reducing the flux at phases around apastron and inferior conjunction.\\
Inverse Compton emission by relativistic Maxwellian pairs in the cold pulsar wind might yield another contribution in this energy band.
The current model does not take this contribution into account explicitly, since the required ultra-relativistic pulsar wind Lorentz factor $\glorentz_p \sim 5000$ cannot be captured by our numerical methods. 
The produced spectra, however, strongly resemble the ones produced by the Maxwellian electrons injected at the shocks in our model \citep[see][]{Takata2014ApJ...790...18T}.
Due to this degeneracy, the latter effectively captures this contribution in our modelling.

\subsubsection{VHE gamma-rays}\label{sec: emission vhe}
\begin{figure}
  \centering
  \includegraphics[width=\linewidth]{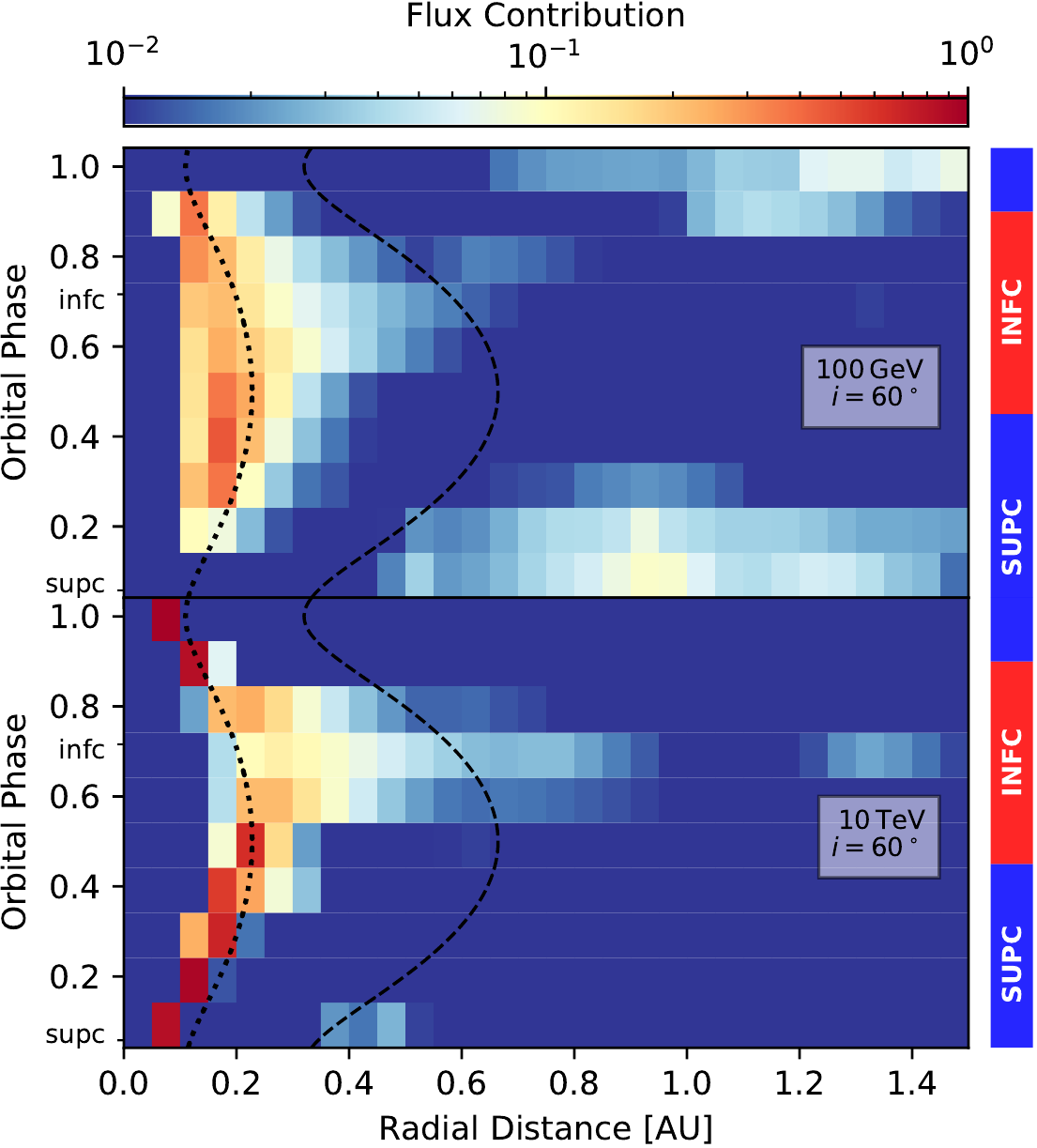}
  \caption{
    Relative contribution to the radiation emitted at a given orbital phase from spherical shells around the system's centre of mass (with a thickness of $0.05\,$ AU) at two different energies.
    For clarity we present the results only for a system inclination of $i=60^\circ$; the resulting maps qualitatively agree with those for an inclination of $i=30^\circ$.
    We terminate the analysis at $r=1.5\,$AU, where the farthest edge of the computational volume is reached.
    Relevant orbital phases are annotated (see also \cref{fig: ls5039 orbit}).
}\label{fig: emission-radial}
\end{figure}In contrast to the HE gamma-ray emission, the flux at VHE is heavily attenuated by $\gamma \gamma$-absorption, introducing an additional line-of-sight dependent modulation.
The observed spectral features and the temporal characteristics are well recovered by our model with a high energy cutoff in excellent agreement with observations and a pronounced anti-correlation to the HE gamma-ray emission, as shown in \cref{fig: emission averaged sed} and the third row in \cref{fig: emission lightcurves}, respectively.
\\
Due to $\gamma \gamma$-absorption, the VHE-flux, especially from the inner region, is strongly reduced for all phases (see \cref{fig: emission averaged sed}).
At INFC, the larger inclination of $i=60^\circ$ is favourable because the absorption is weakened by smaller scattering angles as compared to $i=30^\circ$.
Here, the dominant part of the emission is produced in the middle region with strong relativistic boosting, leading to a hard spectrum as seen in observations.
The increased contribution of this region around inferior conjunction can also be seen more directly in \cref{fig: emission-radial} where we illustrate the relative contribution to the gamma-ray flux as a function of distance from the binary's centre of mass.
\\
In the $100\,$GeV to $1\,$TeV range, where the $\gamma \gamma$-absorption is most pronounced, the flux is underpredicted with respect to observations for SUPC.
At phases around superior conjunction, the emission produced at both the inner and the middle region is almost completely attenuated by absorption, as shown in \cref{fig: emission-radial}, leaving the radiation to be dominantly produced in the far region of the system.
The predicted fluxes, however, are too small to reproduce observations around periastron and superior conjunction (see the third row in \cref{fig: emission lightcurves}), in our simulations.
We found that the region that dominates the flux reaching the observer in the energy range most affected by $\gamma \gamma$-absorption around superior conjunction is extended over a large volume behind the pulsar (see \cref{fig: emission-radial}).
In particular, it extends to the edges of the computational volume, suggesting that a fraction of the emission is lost by particles leaving the domain.
A larger computational domain is required to address this issue, which goes beyond the scope of available computational resources. 
\\
For a system inclination of $i=60^\circ$ we find a sharp peak in the predicted VHE lightcurve at $\phi \simeq 0.55$ (see third row in \cref{fig: emission lightcurves}).
At this phase, the observer's line of sight is aligned with the leading edge of the shocked pulsar wind flow, yielding a drastically increased photon flux due to relativistic boosting (which can also be seen at the bottom of \cref{fig: emission-radial}) that is incompatible with observations.
The trailing edge of the WCR is crossed by the observer between $\phi \simeq 0.8$ and $\phi \simeq 0.9$ in our simulation, showing no peaked emission.

\subsection{Turbulence-induced variability}\label{sec: variability}
To assess the impact of short-term variability on the system's radiative output introduced by turbulence, we evolved the particle distributions presented in \cref{sec:particle results} for more time and recomputed the emission.
After the initial convergence time of $t_0=1.11\,$h, we investigated the particle distributions at $t_n = t_0 + n \cdot \Delta t$, with $\Delta t = 0.14\,$h and $n=0,1,2,3,4$.
This time increment $\Delta t$ is sufficiently longer than the growth-timescale of KH instabilities (see also \cref{sec:fluid results}) for the different solutions to be uncorrelated. 
This procedure was repeated for a range of orbital phases, for which we computed the average and the standard deviation from the five obtained emission solutions and showed the results in \cref{fig: emission lightcurves}.
\\
We found that turbulence introduces variability depending on orbital phase and photon energy with variability levels of several per cent, reaching $\sim 17\%$ for the HE flux around superior conjunction and $\sim 12 \%$ for the VHE flux around inferior conjunction.
Most of the variability is apparent in the inner and middle region, which is expected since these regions are most directly affected by instabilities formed at the contact discontinuity.
Although turbulence introduces visible flux fluctuations, they are still considerably smaller than orbital variations and thus do not dominate the resulting lightcurves.
\\
Lastly, we found that the pulsar-wind shock structure is influenced by turbulence developing at the stellar-wind shock on longer timescales (see \cref{sec:fluid results}).
This has an effect on the variability on orbital timescales and is expected to introduce orbit-to-orbit variations, which will be studied in the future when more computational resources become available.

\subsection{Emission map}\label{sec: emission map}
\begin{figure*}
  \centering
  \includegraphics[width=\linewidth]{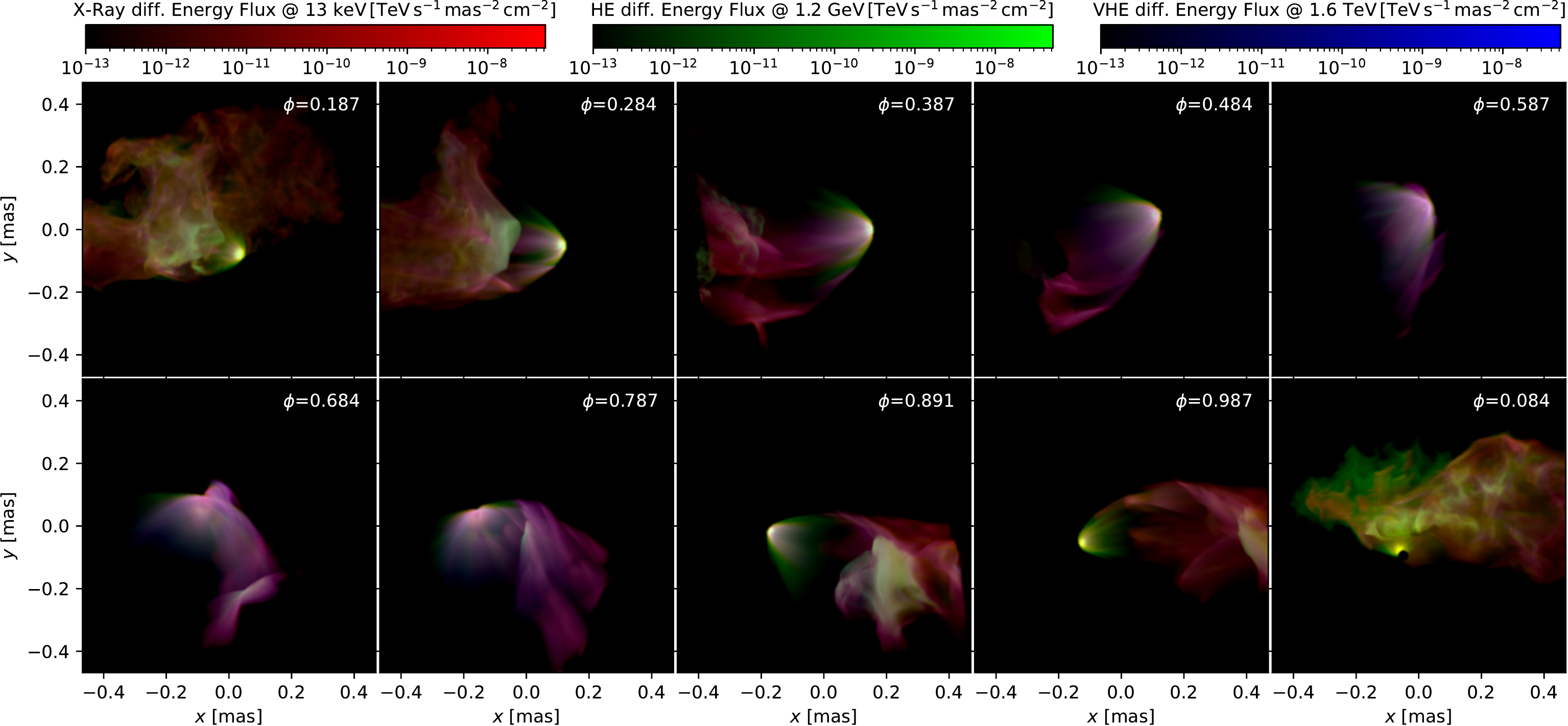}
  \caption{
    Projection of the predicted emission from LS 5039 for different orbital phases as defined with an orbital plane inclination of $i=60^\circ$.
    The differential energy flux is shown for distinct energy bands as indicated by the colour bars.
    For better visualisation, the upper limit of the colour bars has been set two orders of magnitude lower than the maximum HE gamma-ray flux. 
  }\label{fig: emission projections}
\end{figure*}

For better visualisation of the emission produced in LS 5039 we show a composite, false-colour emission map in \cref{fig: emission projections} as it would be seen with a perfect angular resolution at Earth.
The imprint of the fluid-structure is clearly visible, exhibiting the bow shock and the turbulent downstream region behind the pulsar around periastron.
At $\phi \simeq 0.1$ the star is eclipsing parts of the emission region occluding a circular region in the map.
\section{Summary and discussion}\label{sec: summary}
In this work, we present the application of a novel numerical model for gamma-ray binaries \citep{Huber2020-1} to the LS 5039 system.
We specifically choose this system for its broad observational coverage, known orbital parameters and comparison to previous modelling efforts.
\\
Our simulations of the wind interaction in this system show an extended, asymmetric WCR bent by the orbital motion, exhibiting a strong bow-like pulsar-wind shock and a Coriolis shock behind the pulsar.
With our approach, it is, for the first time, possible to consistently account for the complex dynamic shock structure in the particle transport model.
Next to the bow and Coriolis shock, this also includes the reflection shock (formed where the downstream plasma of bow and Coriolis shocks collide) and the secondary shocks, arising in the turbulent downstream medium of the Coriolis shock.
The resulting structures are in agreement with the ones observed in previous simulations \citep[see][]{Bosch-Ramon2015A&A...577A..89B}
, although the Coriolis shock is not formed in the numerical domain at phases around inferior conjunction.
This is caused by the computational domain being to small to capture enough of the leading edge of the WCR to reach the required pressure to terminate the pulsar wind at the respective phases.
In simulations with a larger computational domain, the Coriolis shock is expected to reappear - possibly so close to the pulsar that it would be within the extents of the current computational domain.
\\
We find that the extrema of the fluid's internal energy do coincide with orbital extrema, as assumed by previous studies \citep[see e.g.][]{Dubus2015A&A...581A..27D,Zabalza2013A&A...551A..17Z,Takata2014ApJ...790...18T}.
This affects the resulting particle distributions since most of the models, including the presented one, scale the amount and energetics of accelerated electrons with the available internal fluid energy.
This results in a delay of the electron density extrema of $\Delta \phi \simeq 0.1-0.2$ in our model depending on their energy.
This delay also translates to the production of radiation, however, it is compensated in parts by the onset of relativistic deboosting at superior conjunction (shortly after periastron) in the case of LS 5039.
\\
Our model reproduces the main spectral features of the observed emission from LS 5039 ranging from soft X-rays to VHE gamma-rays, further substantiating a wind-driven interpretation of gamma-ray binaries.
In contrast to previous studies \citep[e.g.][]{Dubus2015A&A...581A..27D} our simulation predicts that the synchrotron emission connects the X-ray emission to the LE gamma-ray emission.
We attribute this result to the different WCR geometry, affecting relativistic boosting, and our improved treatment of the star as an extended photon source, leading to increased $\gamma \gamma$-absorption.
Similar to previous studies \citep[e.g.][]{Dubus2015A&A...581A..27D, Molina2020arXiv200700543M}, we find that a hard spectral index of $s=1.5$ is required for the electron acceleration to describe the observations.
The predicted spectra underestimate the flux in the HE gamma-ray band, presumably lacking a contribution by the magnetospheric emission of the pulsar \citep{Takata2014ApJ...790...18T}.
\\
While the inner region ($<1.2$ orbital separations $d$ from the system's centre of mass) provides the bulk of the emission from X-rays to HE gamma-rays, the middle and far regions ($1.2-3.5 \,d$ and $>3.5\,d$, respectively) are especially relevant for the emission in the VHE band mainly due to relativistic beaming and $\gamma \gamma$-absorption.
Our conclusions contrast the assessment by \citet{Molina2020arXiv200700543M}, who found the far region to be the dominant emitter for all energies and orbital phases.
This directly relates to different assumptions regarding the injection of non-thermal electrons in the system.
While we assume the same fraction of the fluid's internal energy density to be converted to non-thermal electrons at all shocks, \citet{Molina2020arXiv200700543M} assume that a significantly larger fraction of the pulsar's spindown-luminosity is converted at the Coriolis shock than at the bow-shock, which is not the case how we perform our simulation.
\\
Our model predicts significant orbital modulation in all energy bands originating mainly through the anisotropic inverse-Compton process, relativistic boosting, and changing properties of the WCR.
While the predicted HE to VHE gamma-ray lightcurves agree with observations, the simulations fail to reproduce the observed orbital X-ray and LE gamma-ray modulation.
Instead of the observed correlation with VHE gamma-rays, a correlation with the HE band is predicted.
We attribute this to the rather simplistic magnetic field model, i.e. scaling the magnetic energy density with the fluid pressure, which varies strongly across the orbit and is responsible for the dominant modulation in the synchrotron emissivity.
Such a model has been employed in previous studies using a semi-analytical description for the emitting particles \citep[e.g.][]{Barkov2018MNRAS.479.1320B}, but it seems to be increasingly incompatible with our more detailed particle transport model, emphasising the need for a more sophisticated approach.
This could be realised for example by an extension of the presented model to relativistic magneto-hydrodynamics.
With this, it will be possible to take the impact of the non-negligible magnetic field onto the fluid dynamics into account, yielding a more realistic picture of the magnetic field strength and direction, enabling more realistic injection models and the consideration of anisotropic synchrotron emission.
\\
At INFC, the middle region produces the dominant contribution to the VHE gamma-ray spectrum due to relativistic boosting, allowing a hard spectral index to be maintained.
We note that the Coriolis shock is not present in the simulation for most of the corresponding phases, as mentioned in the beginning.
Its presence in larger simulations is expected to reduce the size of the bow shock wings, where this relativistically boosted emission is originating, which might lead to a softening of the spectrum.
\\
At superior conjunction, a significant part of the VHE gamma-ray emission is produced by electrons in the far region of the system due to the weakened impact of $\gamma \gamma$-absorption.
However, emitting particles are lost at the boundaries of the simulation.
We suspect this to be the main reason for the underestimation of the SUPC flux in the $100\,$GeV to $1\,$TeV range and the underestimation of the VHE lightcurve around periastron to superior conjunction.
These issues and the missing Coriolis shock at some orbital phases can be addressed by employing a larger computational volume.
\\
We found that turbulence formed in the wind interaction introduces sub-orbital variability in the systems radiative output on the levels of several per cent on average, reaching up to $\sim 20$ per cent for certain orbital phases and photon energies.
The introduced variations in the flux, however, are still considerably smaller than those on orbital timescales.
On longer timescales, turbulence is expected to introduce orbit-to-orbit variations, which cannot be studied from the single simulated orbit.
This shows the need for further investigations on larger timescales and a larger spatial domain.
\\
Both investigated inclinations of the orbital plane yield fluxes that are more or less consistent with observations depending on the energy band.
The HE gamma-ray lightcurve slightly favours a lower inclination of $i=30^\circ$, due to the amplitude of the modulation, whereas the VHE gamma-ray spectrum prefers the larger inclination of $i=60^\circ$, due to the hard spectral index.
The latter inclination is also favourable because it makes the X-ray lightcurve more consistent with observations by overcoming parts of the strong internal modulation of the synchrotron emissivity due to the increased relativistic boosting.
\\
Although a larger inclination proves to be favourable due to the later decrease of the VHE flux after superior conjunction, the predicted VHE lightcurve at $i=60^\circ$ shows a peak that is incompatible with observations shortly after apastron, arising from relativistic boosting when the leading edge of the shocked pulsar wind flow is crossed by the observer's line of sight \citep[see also][]{Dubus2015A&A...581A..27D}.
We note again, that the amplitude of the VHE peak might be overestimated in the presented work, because of the missing Coriolis shock at the phases of the peak.
\\
Since the employed simulations take the naturally arising asymmetric shape of the WCR into account, the trailing edge is considerably wider as compared to the leading edge when the observer's line-of-sight is crossed, smearing out the effects of relativistic boosting.
This prevents the formation of a second peak in the VHE lightcurve around $\phi \sim 0.85$ seen in previous works with a more simplified prescription of the WCR for higher inclinations \citep[][]{Dubus2015A&A...581A..27D}.
Such a second peak is not visible in related observations.
\\
In this study, we showed the feasibility of a combined fluid and particle-transport simulation for predicting the radiative output of gamma-ray binaries.
Future extensions of this model as discussed shall lead to a better representation of the observations of LS 5039 and will be applied to other well-observed gamma-ray binaries.

\begin{acknowledgements}
   The computational results presented in this paper have been achieved (in parts) using the research infrastructure of the Institute for Astro- and Particle Physics at the University of Innsbruck, the LEO HPC infrastructure of the University of Innsbruck, the MACH2 Interuniversity Shared Memory Supercomputer and PRACE resources.
   We acknowledge PRACE for awarding us access to Joliot-Curie at GENCI@CEA, France.
   This research made use of Cronos \citep{Kissmann2018ApJS..236...53K}; GNU Scientific Library (GSL) \citep{galassi2018scientific}; matplotlib, a Python library for publication quality graphics \citep{Hunter2007}; and NumPy \citep{van2011numpy}.
   We thank the anonymous referee for the thoughtful comments and suggestions that allowed us to improve our manuscript.
\end{acknowledgements}

\bibliographystyle{aa}
\bibliography{references}

\begin{appendix}
   \section{Supplementary material}
\begin{figure*}
  \centering
  \includegraphics[width=\linewidth]{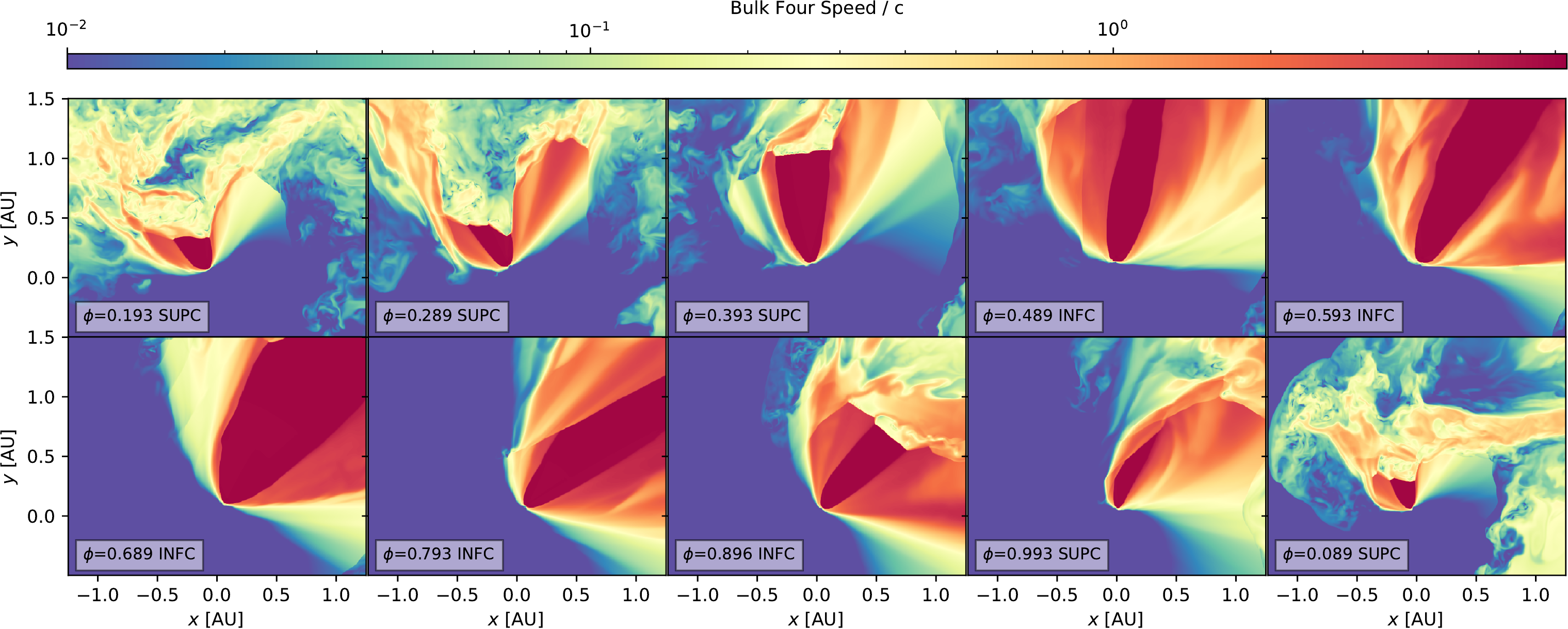}
  \caption{
    The fluid's four-speed in analogy to \cref{fig: results fluid density}.
  } \label{fig: results fluid four speed}
\end{figure*}\begin{figure*}
  \centering
  \includegraphics[width=\linewidth]{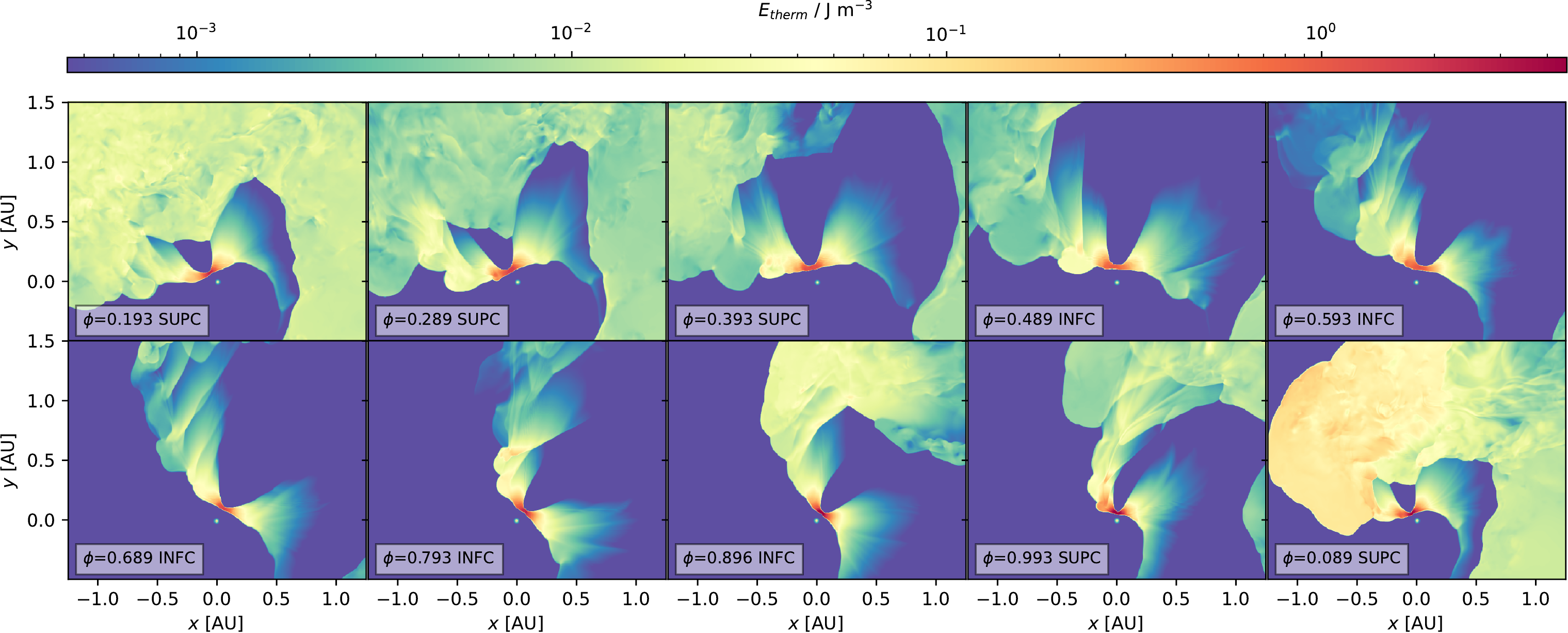}
  \caption{
    The fluid's thermal energy in analogy to \cref{fig: results fluid density}.
  } \label{fig: results fluid thermal energy}
\end{figure*}\begin{figure*}
  \centering
  \includegraphics[width=\linewidth]{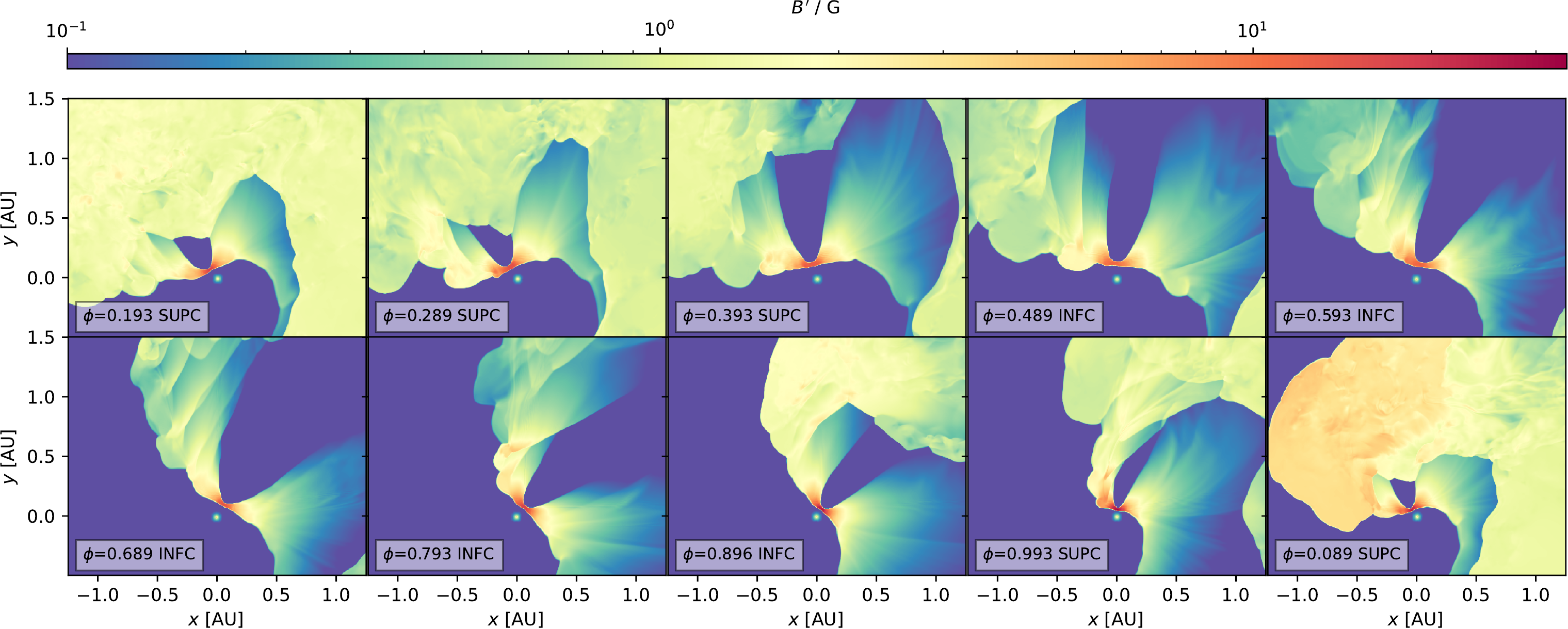}
  \caption{
    The magnetic field in the fluid frame in analogy to \cref{fig: results fluid density}.
  } \label{fig: fluid magnetic field}
\end{figure*}In this section we provide additional visualisation for the fluid quantities, that have been omitted in the main text.
In \cref{fig: results fluid four speed}, \cref{fig: results fluid thermal energy} and \cref{fig: fluid magnetic field} we show the fluid's four-velocity, its thermal energy density and the magnetic field strength in the fluid frame, respectively. \end{appendix}

\end{document}